\documentclass[aps,twocolumn,floats,superscriptaddress,longbibliography,nofootinbib,nopacs]{revtex4-1}
\usepackage{braket}
\usepackage{booktabs}
\usepackage{graphicx}
\usepackage{dcolumn}
\usepackage{bm}
\usepackage{color}
\usepackage{amsmath}
\usepackage{amssymb}
\usepackage{hyperref}
\usepackage{verbatim}

\newcommand{\h}{\mathit{\boldsymbol{h}}}
\newcommand{\X}{\mathit{\boldsymbol{x}}}

\newcommand{\cl}{\mathrm{cl}}

\newcommand{\gf}{G}

\newcommand{\beq}{\begin{equation}}
\newcommand{\eeq}{\end{equation}}
\newcommand{\ba}{\begin{array}}
\newcommand{\ea}{\end{array}}
\newcommand{\bea}{\begin{eqnarray}}
\newcommand{\eea}{\end{eqnarray}}

\begin{document}

\title{Tunneling in projective quantum Monte Carlo simulations with guiding wave functions }

\author{T.~Parolini}
\affiliation{SISSA - International School for Advanced Studies, I-34136 Trieste, Italy}
\affiliation{The Abdus Salam International Centre for Theoretical Physics, I-34151 Trieste, Italy}
\affiliation{Istituto Nazionale di Fisica Nucleare, Sezione di Trieste, I-34136 Trieste, Italy}

\author{E.~M.~Inack}
\affiliation{SISSA - International School for Advanced Studies, I-34136 Trieste, Italy}
\affiliation{The Abdus Salam International Centre for Theoretical Physics, I-34151 Trieste, Italy}
\affiliation{Perimeter Institute for Theoretical Physics, Waterloo, Ontario, Canada N2L 2Y5}
\affiliation{INFN, Sezione di Trieste, I-34136 Trieste, Italy}

\author{G.~Giudici}
\affiliation{SISSA - International School for Advanced Studies, I-34136 Trieste, Italy}
\affiliation{The Abdus Salam International Centre for Theoretical Physics, I-34151 Trieste, Italy}
\affiliation{Istituto Nazionale di Fisica Nucleare, Sezione di Trieste, I-34136 Trieste, Italy} 

\author{S.~Pilati}
\affiliation{School of Science and Technology, Physics Division, Universit{\`a}  di Camerino, 62032 Camerino (MC), Italy}

\begin{abstract}
Quantum tunneling is a valuable resource exploited by quantum annealers to solve complex optimization problems. 
Tunneling events also occur during projective quantum Monte Carlo (PQMC) simulations, and in a class of problems characterized by a double-well energy landscape their rate was found to scale linearly with the first energy gap, i.e.,
even more favorably than in physical quantum annealers, where the rate scales with the gap squared.
Here we investigate how a guiding wave function---which is essential to make many-body PQMC simulations computationally feasible---affects the tunneling rate. 
The chosen test beds are a continuous-space double-well problem, the ferromagnetic quantum Ising chain, and the recently introduced shamrock model. 
As guiding wave function, we consider an approximate Boltzmann-type ansatz, the numerically exact ground state of the double-well model, and a neural-network wave function based on a Boltzmann machine.
Remarkably, for each ansatz we find the same asymptotic linear scaling of the tunneling rate that was previously found in the PQMC simulations performed without a guiding wave function.
We also provide a semiclassical theory for the double well with exact guiding wave function that explains the observed linear scaling.
These findings suggest that PQMC simulations guided by an accurate ansatz represent a valuable benchmark for physical quantum annealers and a potentially competitive quantum-inspired optimization technique.
\end{abstract}

\maketitle

\section{Introduction}
\label{secintro}
Quantum annealers are special purpose adiabatic quantum computers designed to solve complex optimization problems~\cite{boixo2014evidence}. 
Compared to alternative classical optimization algorithms, chiefly simulated annealing, they can additionally exploit quantum tunneling to cross energy barriers and reach lower energy solutions~\cite{santorotheory,nishimoriising,albash2018adiabatic,hauke2019perspectives}.
Their dominant bottlenecks are the small energy gaps associated to avoided level crossings. Such small gaps typically occur in disordered systems when two well-separated competing states are connected by a tunneling process. This scenario frequently happens in the glassy phases that characterize typical hard optimization problems.

Simulating the real-time dynamics of quantum annealers, e.g., to identify classes of problems where they might outperform classical optimization methods, is possible only for relatively small systems (say, around 30 spins).
Quantum Monte Carlo (QMC) simulations have emerged as a useful alternative tool to simulate the quantum annealers' behavior 
in configurations where the sign problem does not occur~\cite{Finnila_CPL94,santoroPIMC,santoroGFMC, troyerheim,boixo2014evidence,albash2015reexamining,mbeng2019dynamics}. This is the case, e.g, of the devices currently commercialized by \textit{D}-wave systems (see, e.g.,~\cite{johnson2011quantum,boixo2013experimental,boixo2014evidence,ronnowspeedup2014,lantingentanglement2014}.
In particular, path-integral Monte Carlo (PIMC)~\cite{RevModPhys.67.279} and projective QMC (PQMC)~\cite{foulkes2001quantum} algorithms have been adopted, beside other techniques such as the stochastic series expansion algorithm~\cite{liu2015,liu2013}.
Tunneling events also occur during QMC simulations, similarly to what happens during the quantum annealers' dynamics~\cite{boixo2016computational,denchevtunneling2016}. 
In various problems characterized by a double-well energy landscape, the tunneling rate of finite-temperature PIMC simulations was found to scale with the system size, or with the height of the energy barrier, as the square of the first energy gap~\cite{brady2016quantum,isakovtunneling,mazzolaquantumchemistry}. 
This is the same scaling predicted by the theory of incoherent quantum tunneling~\cite{weiss1987incoherent}, and it is also the scaling of the inverse of the annealing time required by a coherent quantum annealer to avoid diabatic transitions~\cite{farhi2000quantum}.

References~\cite{isakovtunneling,mazzolaquantumchemistry,PhysRevA.96.042330} explained these results using a semiclassical theory of instantons in PIMC simulations.
In the case of PQMC algorithms, the tunneling rate was found to scale linearly with the gap, providing a quadratic speedup compared to the expected behavior of a quantum annealer~\cite{inack2}.\footnote{A linear scaling was identified also for PIMC simulations performed with open boundary conditions in the inverse-temperature direction~\cite{isakovtunneling,mazzolaquantumchemistry}. However, significant deviations have later been discussed~\cite{PhysRevA.96.042330}. Furthermore, the computational cost of zero-temperature simulations based on open-boundary PIMC algorithms has not been analyzed in detail.}
Furthermore, the PQMC algorithms displayed this speedup even in the so-called shamrock model~\cite{inack2}, where frustrated interactions cause an exponential slowdown of the finite-temperature PIMC dynamics~\cite{aminshamrock}.
These findings suggest that PQMC simulations constitute a relevant benchmark for physical quantum annealers and a competitive quantum-inspired classical optimization algorithm~\cite{inack,crosson2016simulated,jordanDMC}. 
In fact, they have recently been employed to obtain better solutions in optimization problems relevant for medical research, specifically, for pulse-sequence optimization in magnetic resonance fingerprinting~\cite{Microsoft}.
This further highlights the importance of exhaustively characterizing their tunneling dynamics.\\

The tunneling-time studies mentioned above have considered PQMC algorithms implemented without a guiding wave function (GWF). 
The GWF, usually a variational ansatz that closely approximates the ground state, guides the PQMC simulation towards the relevant regions of the configuration space. This  improves the algorithm's accuracy and efficiency~\cite{foulkes2001quantum}. In fact, without a sufficiently accurate GWF, the computational cost of PQMC simulations increases exponentially with the system size~\cite{nemec,boninsegnimoroni,inack2}.
In principle, one might expect the GWF to significantly impact the tunneling dynamics, since it alters both the sampling algorithm and the probability distribution sampled at equilibrium.
In this paper, we analyze whether the GWF does indeed affect the tunneling time in PQMC simulations, and if it does, to what extent.
As test beds, we consider a one-dimensional continuous-space Hamiltonian describing a quantum particle in a double well, the ferromagnetic quantum Ising chain, and the shamrock model. Notice that, in the ferromagnetic phase, also the Ising-type models can be described by an effective double-well potential, with the two polarized states with opposite magnetizations representing the two competing potential minima.
We consider different kinds of GWFs, including a Boltzmann-type ansatz that mimics the equilibrium distribution of a classical statistical ensemble and, for the continuous-space model, the numerically exact representation of the ground state. For the quantum Ising chain, we also consider an ansatz that mimics the structure of a generative artificial neural network~\cite{carleotroyer}, specifically an unrestricted Boltzmann machine~\cite{inack3} or, in a different jargon, a shadow wave function~\cite{reatto1988shadow,shadowvitiello}. 
Remarkably, for all GWFs we consider, we find the same linear scaling (to leading exponential order) of the tunneling rate with the gap as previously found in PQMC simulation performed without GWF. The choice of the GWF only affects the prefactor. 
We also provide a semiclassical theory based on the Wentzel--Kramers--Brillouin (WKB) approximation, valid for PQMC simulations of the double well with exact GWF, that explains the observed linear scaling.

The rest of the paper is organized as follows: in Sec.~\ref{secdoublwell} we present the PQMC algorithm, implemented with and without GWF, and the analysis of the tunneling time for the continuous-space double-well model. 
In the same section, the WKB semiclassical theory of the PQMC tunneling time is reported.
The algorithm and the tunneling-time analysis for the quantum Ising chain and for the shamrock model are presented in Sec.~\ref{secQIC}.
Our conclusions and the outlook are reported in Sec.~\ref{secconclusions}.
 
\section{Tunneling time in continuous-space double-well systems}
\label{secdoublwell}
\begin{figure}
\includegraphics[width=\columnwidth]{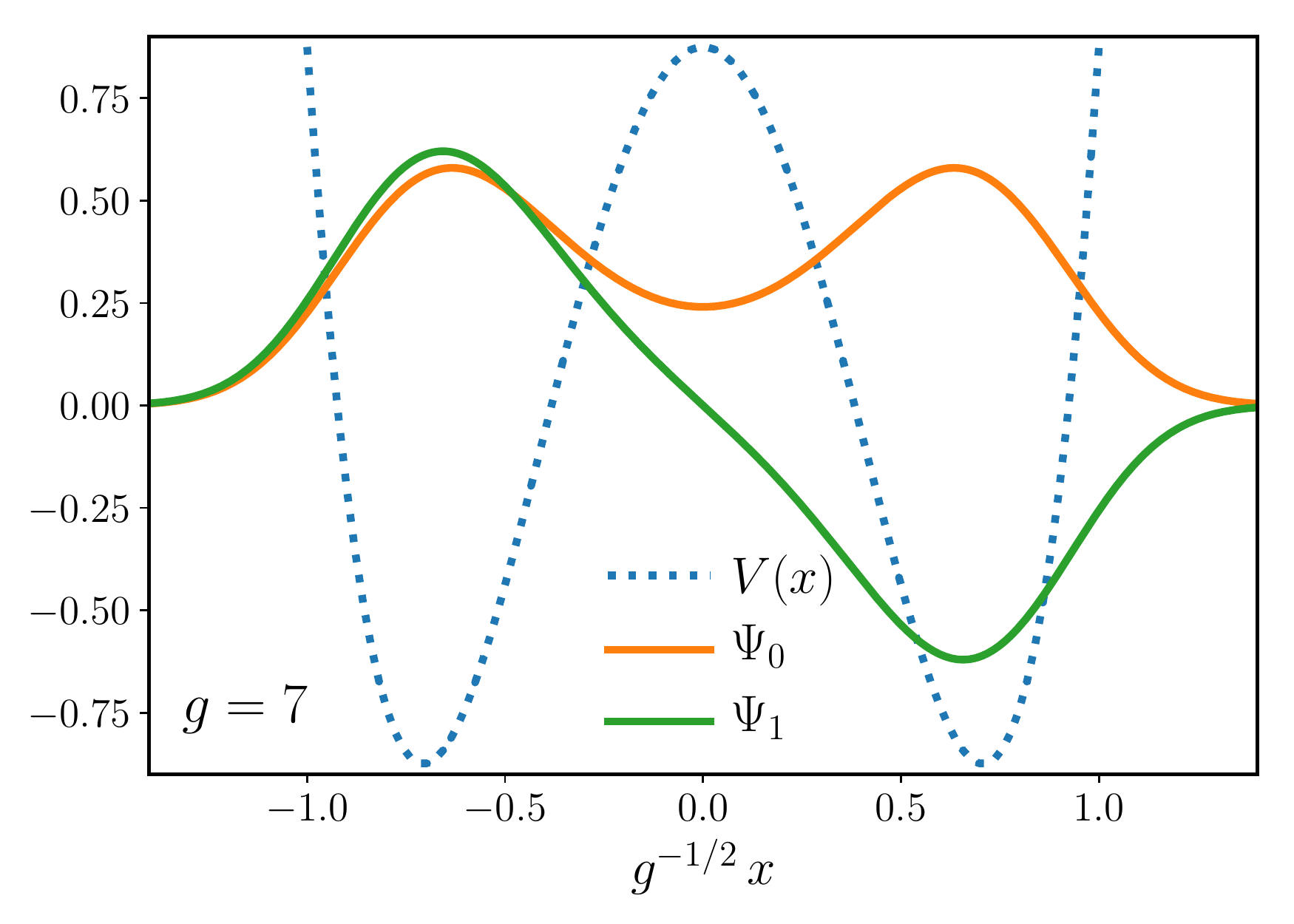}
\caption{(color online). Profile of the quartic double-well potential Eq.~\eqref{DW-potential} (dotted blue line), shifted for better comparison with the wave functions of the ground state $\Psi_0(x)$ and of the first excited state $\Psi_1(x)$ (orange and green solid lines, respectively).}
\label{fig1}
\end{figure}

In this section, we consider a quantum particle in one spatial dimension, described by the following continuous-space Hamiltonian:
\begin{equation}\label{double-well-H}
\hat{H} = -\frac{1}{2}\frac{\mathrm{d}^2}{\mathrm{d}x^2} + V(x),
\end{equation}
with the quartic double-well potential
\begin{equation}\label{DW-potential}
V(x) = \frac{x^4}{g} - x^2.
\end{equation}
The profile of \( V(x) \) is visualized in Fig.~\ref{fig1}, together with the corresponding ground-state wave function \( \Psi_0(x) \) and the first excited state \( \Psi_1(x) \).
Units are chosen so that \( \hbar/m=q=1 \), where \( \hbar \) is the reduced Planck constant, \( m \) is the particle mass and \( q = \frac{1}{2}\sqrt{V^{\prime\prime}(x_{L,R})} \) fixes the curvature at the bottom of the well. Here \( x_{L,R} = \mp \sqrt{g/2} \) are the minimum points of \( V(x) \).
The dimensionless parameter \( g \) controls the height of the barrier separating the two wells, \( V(0) - V(x_L) \propto g \), and the distance between the minimum points, \( x_R-x_L \propto \sqrt{g} \).
For large \( g \), the barrier height increases relative to the wells' separation and the energy spectrum becomes  doubly degenerate, corresponding to the two wells being asymptotically independent. 
For large but finite \( g \), the two wells are connected by tunneling processes. These processes lift the degeneracies, leading to small tunneling gaps between the energy levels. 
An approximate expression for the first energy gap \( \Delta = E_1-E_0 \) between the ground state energy $E_0$ and the first excited level $E_1$, valid in the large-\( g \) regime, can be determined via the WKB semiclassical theory. One obtains~\cite{garg}
\begin{equation}\label{WKBgap}
\Delta = 8\sqrt{\frac{g}{\pi}} \exp\left(-\frac{2g}{3}\right).
\end{equation}
In an isolated double well, a quantum particle initially prepared in one of the two states
\begin{equation}\label{psiLR}
	\Psi_{L,R}(x) = \frac{\Psi_0(x)\pm \Psi_1(x)}{\sqrt{2}},
\end{equation}
which are localized in the left and right wells, respectively, performs coherent periodic oscillations between the two states at a rate proportional to \( \Delta \). 
As a reference, it is worth mentioning that, instead, a quantum particle coupled to a thermal bath and subjected to a double-well potential would undergo incoherent tunneling at a rate proportional to \( \Delta^2 \)~\cite{weiss1987incoherent}.

\subsection{Diffusion Monte Carlo algorithm}\label{subs:DMC}
We are interested in comparing the tunneling dynamics generated by PQMC simulations to the quantum tunneling time, and also in identifying their relation with the energy gap \( \Delta \).  Specifically, we consider simulations performed with the Diffusion Monte Carlo (DMC) method, which belongs to the family of PQMC algorithms for continuous-space systems.
The DMC algorithm aims at projecting out the ground state by evolving the imaginary-time Schr\"odinger equation for the time-dependent wave function \( \Psi (x,t) \). In one dimension, this equation reads
\begin{equation}\label{DMC-NoIS-PDE}
	\frac{\partial\Psi(x,t)}{\partial t} = \frac{1}{2}\frac{\partial^2 \Psi(x,t) }{\partial x^2} - \left[V(x)-E_T\right]\Psi(x,t).
\end{equation}
Here, \( t \) is the imaginary time and \( E_T \) is an energy threshold introduced to stabilize the numerics, as explained later.  
Following the seminal work of Ref.~\cite{andersonJB}, one can exploit the analogy between Eq.~\eqref{DMC-NoIS-PDE} and a modified diffusion equation, i.e., one including a source/sink process corresponding to the second term on the right-hand side, to implement a stochastic simulation. 
Due to the norm-nonconserving nature of this source/sink term, the simulation has to evolve a large population of equivalent instances of system configurations \( \{x_i\} \), in jargon called walkers, subject to diffusion and to a killing and replication process called branching.
The sampling algorithm is dictated by a short-time approximation of the Green function of Eq.~\eqref{double-well-H}. This is assumed to apply for a short imaginary-time step \( \tau \). Long imaginary times \( t=N_{\mathrm{DMC}}\tau \) can be reached by iterating many DMC steps \( N_{\mathrm{DMC}} \), each corresponding to a time step \( \tau \). In the long imaginary time limit, the walkers sample configurations with a probability distribution proportional to \( \Psi(x,t\rightarrow\infty)\propto \Psi_0(x) \).
Notice that \( \Psi_0(x) \) is assumed to be a real and nonnegative function. This is legitimate in the case of stoquastic Hamiltonians, i.e., those not affected by the negative-sign problem~\cite{bravyi2008complexity,bravyi2009complexity}.
We adopt the linear Trotter approximation for the short-time Green function. This leads to the following sampling algorithm. After initializing, say, $N_w$ walkers in some configurations \( \{x_i\} \), at every DMC step one applies to each walker \( i \)  two processes: first, the configuration update \( x^\prime_i = x_i + \sqrt{\tau}\delta_i \), where \( \delta_i \) is sampled from a zero-mean, unit-variance Gaussian distribution; then, one samples an integer number \( s_i = \lfloor{w_i+r_i}\rfloor \), where \( w_i = \exp\left[{-\tau\left(V(x^\prime_i)-E_T\right)}\right] \) is the walker weight and \( r_i\in(0,1) \) is a uniform random variable, and creates \( s_i \) copies of the walker to be included in the population for the next time step.
The size of the walker population can be tuned close to a desired target value \( N_w \) by appropriately controlling \( E_T \). For this control, we adopt the textbook recipe of Ref.~\cite{thijssen}. Therein, the interested readers will also find a more pedagogical description of the DMC algorithm.
The systematic bias due to the Trotter approximation can be eliminated via zero time-step extrapolation. A more subtle bias might originate also from the finite walker population \( N_w \)~\cite{nemec,boninsegnimoroni,pollet2018stochastic}. Indeed, it has been shown that in many-body systems, in order to keep this bias below a chosen small threshold, \( N_w \) has to exponentially increase with the system size~\cite{inack2}.
In order to reduce or eliminate this systematic bias and to reduce the statistical fluctuations, it is standard practice to adopt an importance sampling approach using a guiding wave function \( \Psi_G(x) \)~\cite{foulkes2001quantum}. Typically, \( \Psi_G(x) \) is a parametrized variational ansatz, whose optimal parameters are determined via energy expectation-value minimization.
Hence, one evolves the modified imaginary-time Schr\"odinger equation for the product \( \rho(x,t)=\Psi(x,t)\Psi_G(x) \), which takes the form of a Fokker--Planck type equation with an additional source/sink term. This equation reads:
\begin{equation}\label{DMC-IS-PDE}
	\frac{\partial\rho}{\partial t} = \frac{1}{2}\frac{\partial^2\rho}{\partial x^2} + \frac{\partial}{\partial x}\left[\tilde{V}^\prime(x)\rho\right] - \left[E_L(x)-E_T\right]\rho,
\end{equation}
where we wrote \( \rho \) for \( \rho(x,t) \) to simplify the notation, the local energy is
\begin{equation}
E_L(x) = V(x) - \frac{1}{2\Psi_G(x)}\frac{\mathrm{d}^2 \Psi_G(x)}{\mathrm{d}x^2},
\end{equation}
and we introduced the effective potential
\begin{equation}\label{Veff}
	\tilde{V}(x) = -\ln\Psi_G(x).
\end{equation}
Adopting, again, a linear approximation for the Green function, the sampling algorithm is modified as follows: the configuration update includes, beside the Gaussian random term, a deterministic displacement computed as \( \tau \frac{\mathrm{d}}{\mathrm{d}x}\ln\Psi_G(x) \); in the branching process, the potential \( V(x) \)  is replaced by \( E_L(x) \) for the computation of the walker weight \( w_i \).
In the long imaginary-time limit, the walkers sample a probability distribution proportional to \( \rho(x,t\rightarrow \infty)=\Psi_0(x)\Psi_G(x) \).
One should notice that \( E_L(x) \) is a constant function if \( \Psi_G(x) \) is an exact eigenstate of the Hamiltonian. This completely suppresses the fluctuations of the random walker population, eliminating the finite-\( N_w \) bias. In fact, the algorithm's accuracy and efficiency are significantly improved even when \( \Psi_G(x) \) is, albeit not exact, a reasonably good approximation of the ground-state wave function.

\subsection{Tunneling time in Diffusion Monte Carlo simulations}
Our goal is to analyze the relation between the DMC tunneling time \( \xi \), namely the imaginary time required by the walkers to leak from one well to the other, and the physical tunneling time of the real-time dynamics. More precisely, we are interested in the scaling relation between \( \xi \) and the inverse energy gap \( \Delta^{-1} \).
We measure \( \xi \) with a protocol analogous to the one adopted in Refs.~\cite{isakovtunneling,mazzolaquantumchemistry} for PIMC simulations. 
All walkers are initially set at the bottom of the left well \( x=x_L \). The DMC simulation is run until a percentage \( p \) of the instantaneous walker population overcomes a position threshold in the right well \( x_\textrm{th} \ge 0 \). 
The final imaginary time is recorded, and the process is repeated approximately 300 times to accumulate statistics. The fluctuations of the final imaginary times turn out to be approximately normally distributed, and we take the average and its standard deviation as the definition of \( \xi \) and of its error bar, respectively.
In the simulations reported here, the threshold is set at \( x_\textrm{th} = x_R/2 \) and the walker percentage at \( p = 25\% \). A careful analysis shows that the asymptotic scaling  of  \( \xi \) is independent of this specific choice up to a constant prefactor.
Furthermore, the chosen target walker population \( N_w\approx10^4 \) is large enough, and the chosen time step small enough (e.g., \( \tau = 0.007 \)  for \( 1/\Delta > 70 \)), to eliminate any significant effect on \( \xi \).\\

\begin{figure}
\includegraphics[width=\columnwidth]{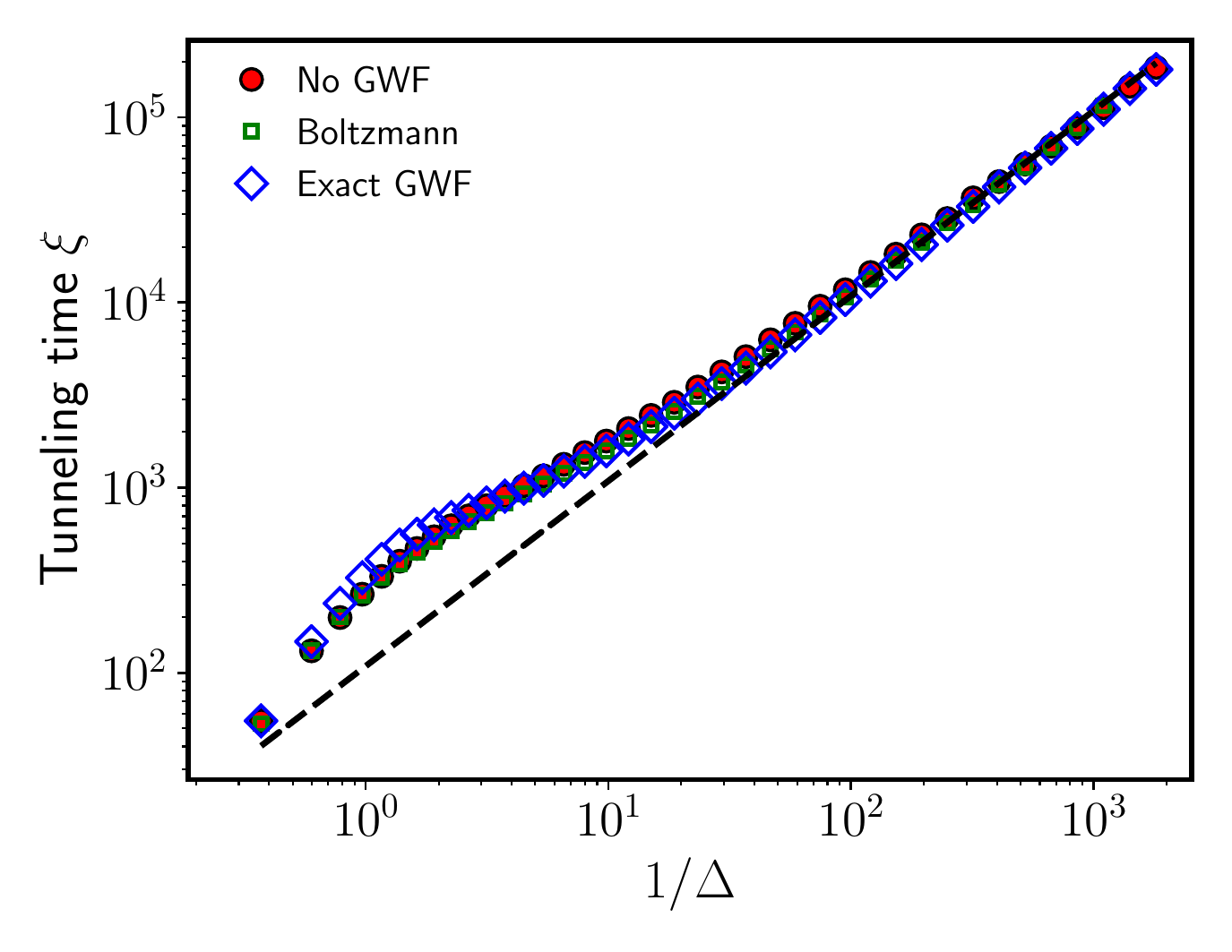}
\caption{(color online). DMC tunneling time \( \xi \) for the quartic double-well potential~\eqref{DW-potential} as a function of the inverse energy gap \( \Delta^{-1} \). Three different DMC protocols are shown: the simple DMC algorithm without a GWF (red circles), the DMC algorithm guided by a Boltzmann ansatz (green empty squares), and the one guided by the numerically-exact representation of the ground state $\Psi_0(x)$ (blue empty diamonds). The dashed line represents the scaling \( \xi \propto \Delta^{-1} \). Here and in all plots, if not visible, the error bars are smaller than the symbol size.}
\label{fig:tunn_time_DW}
\end{figure}

First, the DMC algorithm is run without a guiding wave function (GWF). We measure the tunneling time \( \xi \) for different barrier heights, tuned by varying the parameter \( g \). In Fig.~\ref{fig:tunn_time_DW}, \( \xi \) is plotted as a function of the inverse energy gaps \( \Delta^{-1} \), which we compute for the different \( g \) values we consider using a standard finite-difference  method. The discretization is fine enough to ensure that there is no sizable finite-precision effect.
In the large-\( g \) regime, corresponding to large \( \Delta^{-1} \), the tunneling times approach the scaling law \( \xi \propto \Delta^{-1} \).
This scaling relation has previously been identified in Ref.~\cite{inack2} in PQMC simulations of Ising-type models, again performed without a GWF.
Next, we run DMC simulations with a GWF. First, as GWF we consider the approximate Boltzmann ansatz \( \Psi_G(x) = \exp\left[{-\beta V(x)}\right] \).
The fictitious inverse temperature \( \beta \) is fixed by minimizing the variational energy estimate. Second, we consider a numerical representation of the exact ground-state wave function, i.e., we set \( \Psi_G(x)=\Psi_0(x) \). The ground-state wave function is obtained via the finite-difference technique.  Notice that this ansatz represents the optimal GWF for equilibrium simulations.
In Fig.~\ref{fig:tunn_time_DW}, the tunneling times obtained with these two GWFs are compared with the results obtained without GWF.
Remarkably, for large \( g \), the same linear relation between \( \xi \) and \( \Delta^{-1} \) is approached. By fitting the three datasets in the large-\( g \) regime with the function \( \xi(\Delta)=\alpha\Delta^{-b} \), where \( \alpha \) and \( b \) are the fitting parameters, we obtain the values reported in Table~\ref{tab:fit}. In all three cases, the exponent \( b \) is consistent with the linear relation corresponding to \( b=1 \). The choice of GWF only affects the prefactor \( \alpha \), though for this model the variations are small enough to be masked by statistical uncertainties.
These findings indicate that the GWF does not affect the leading scaling relation between tunneling time and inverse energy gap. This is a surprising results, given that introducing the GWF affects both the sampling algorithm and the equilibrium probability distribution of the DMC simulation.
A rough explanation can be conjectured by considering the competition between two effects originating from the introduction of the GWF. The first is due to the deterministic drift, which pushes walkers away from the potential barrier, inhibiting inter-well crossings. The second is the smoothing out of the weight reduction that occurs when walkers encounter a bump in the potential; this effect reduces the probability of those walkers being eliminated from the population. Our numerical results indicate that these two effects tend to compensate. 
A more formal explanation of the linear relation between \( \xi \) and \( \Delta^{-1} \) is given in the next subsection.

The double-well potential \eqref{DW-potential} is characterized by a specific functional relation between the barrier's height and width. In order to verify that our findings do not rely on this particular choice, we introduce an adjustable parameter that allows us to vary the width of the barrier independently of its height. Similarly to Ref.~\cite{mazzolaquantumchemistry}, we consider the potential
\begin{equation}\label{DW-plateau}
U(x) = \frac{\left(|x|-x_0\right)_+^4}{g} - \left(|x|-x_0\right)_+^2,
\end{equation}
where \( f(x)_+ \equiv \max\{0,f(x)\} \) and \( x_0 \ge 0 \). This potential features a \emph{plateau} of width \( 2x_0 \) around the origin, and reduces to \( V(x) \) for \( x_0 = 0 \).
In our study, the barrier height is kept constant by fixing \( g \) to some value (we use \( g = 8 \)), while \( x_0 \) is increased in the interval \( x_0 \in [0,2] \). This has the effect of reducing the gap \( \Delta \) as well as the tunneling rate \( 1/\xi \). The GWF chosen for this study is the numerically exact ground-state wave function.
Once again, a linear relation is found between \( \xi \) and \( \Delta^{-1} \), and fitting the dataset with the power law \( \xi(\Delta) = \alpha\Delta^{-b} \) in the small-gap regime yields the values \( \alpha = 23(1) \) and \( b = 0.993(7) \) for the parameters.

\begin{table}
\begin{tabular}{p{7em}p{5em}l}
DMC & \( \alpha \) & \( b \)\\\cline{1-3}\\[-2ex]
\textsf{No GWF} & 109(11) & 0.99(2)\\
\textsf{Boltzmann} & 98(16) & 1.01(3)\\
\textsf{Exact GWF} & 112(8) & 0.99(1)\\
\end{tabular}
\caption{Fitting parameters \( \alpha \) and \( b \), describing the small-gap behavior of the DMC tunneling time \( \xi \) in the double-well potential~\eqref{DW-potential} according to the fitting function \( \xi(\Delta) = \alpha \Delta^{-b} \). For each protocol, the five rightmost data points shown in Fig.~\ref{fig:tunn_time_DW} are included in the fit.}
\label{tab:fit}
\end{table}

\subsection{Semiclassical theory of the DMC tunneling dynamics}\label{subs:semiclassical}
We present here a semiclassical theory to explain and generalize our findings.
As mentioned above, the DMC algorithm with GWF is described  by a Fokker--Planck equation, Eq.~\eqref{DMC-IS-PDE}, containing both the usual drift and diffusion terms, and an additional norm-nonpreserving term corresponding to the branching process. When the GWF coincides with the exact ground-state wave function \( \Psi_G(x) = \Psi_0(x) \), this term can be eliminated and branching does not occur. The resulting equation reads:
\begin{equation}\label{DMC-PDE}
	\frac{\partial\rho(x,t)}{\partial t} = \frac{1}{2}\frac{\partial^2\rho(x,t)}{\partial x^2} + \frac{\partial}{\partial x}\left[\tilde{V}^\prime(x)\rho(x,t)\right].
\end{equation}
This equation describes the stochastic dynamics of a classical particle with distribution \( \rho(x,t) \) subject to the effective potential \( \tilde{V}(x) \) defined in Eq.~\eqref{Veff}. The tunneling time corresponding to this stochastic dynamics can then be identified (to exponential accuracy) with the activation time needed for this classical ensemble to overcome the effective barrier \( \Delta\tilde{V} \equiv \tilde{V}(0) - \tilde{V}(x_\textrm{min}) \), with initial conditions \( \rho(x,0) = \delta(x-x_L) \) (the normalization can be set to one since the norm of \( \rho(x,t) \) is conserved). Here, \( x_\textrm{min} \) indicates the (left) minimum point of \( \tilde{V}(x) \), in general different from \( x_L \), while \( x = 0 \) is its local maximum point, as follows from the form of \( \Psi_0(x) \).
The computation of the classical activation over a potential barrier is known as Kramers problem~\cite{kramers,kamenev}. The corresponding  activation time reads
\begin{equation}\label{tau-act}
	\tau_\textrm{act} = \frac{2\pi}{\sqrt{\tilde{V}^{\prime\prime}(x_\textrm{min})\left|\tilde{V}^{\prime\prime}(0)\right|}}\exp\!\big(2\Delta\tilde{V}\big).
\end{equation}
The large-\( g \) scaling of \( \tau_\textrm{act} \) is dominated  by the exponential function in Eq.~\eqref{tau-act}, so we will focus on that term only.
Substituting  Eq.~\eqref{Veff} into Eq.~\eqref{tau-act}, one obtains
\begin{equation}\label{tau-psi}
	\tau_\textrm{act} \sim \left(\frac{\Psi_0(0)}{\Psi_0(x_\textrm{min})}\right)^{-2} \sim \Psi_0(0)^{-2}.
\end{equation}
Here, we used the fact that \( \Psi_0(x_\textrm{min}) \) cannot be exponentially small in \( g \) if the ground-state wave function is to be normalized to unity (notice that, by definition, \( \Psi_0(x) \) achieves its global maximum at \( x_\textrm{min} \), and its amplitude is exponentially small outside of a region of width \( O(\sqrt{g}) \)).
The value of \( \Psi_0(0) \) can be estimated using WKB theory, which gives \( \Psi_0(0) \sim \exp(-g/3) \)~\cite{muller-kirsten}.
Substituting this into Eq.~\eqref{tau-psi}, and combining with the WKB estimate of the gap, Eq.~\eqref{WKBgap}, we finally get
\begin{equation}\label{tau-delta}
	\xi \sim \tau_\textrm{act} \sim \frac{1}{\Delta},
\end{equation}
where the symbol \( \sim \) denotes asymptotic equal scaling for \( g \rightarrow \infty \), up to subexponential corrections.\\
The above argument can actually be generalized to double well--type potentials other than the quartic double well defined in Eq.~\eqref{DW-potential}. In fact, Eq.~\eqref{tau-psi} can be derived from Eq.~\eqref{DMC-PDE} under quite general assumptions on \( V(x) \), as long as a sufficiently accurate approximant of the ground-state wave function is available to use as a GWF, such that the effect of branching may be neglected. 
Moreover, for a generic double-well potential the ground-state gap in the large-barrier limit can be expressed as \cite{garg,landaulifshitz}: 
\begin{equation}\label{LLgap}
\Delta \propto \Psi_R(0) \Psi_R^\prime(0),
\end{equation}
where \( \Psi_R(x) \) is the ground-state wave function of a \emph{single} well, as defined in Eq.~\eqref{psiLR}.
The above relation holds in the only hypothesis that \( \Psi_R(x) \) is asymptotically localized on \( x > 0 \), such that the condition \( \int_{-\infty}^0\left[\Psi_R(x)\Psi_R(-x) - \Psi_R(x)^2\right]\mathrm{d}x \ll 1 \) is fulfilled for large enough \( g \).
We can now resort to WKB theory, which gives the following expression for \( \Psi_R(x) \) (to exponential accuracy):
\begin{equation}
\Psi_R(x) = \exp \left(-\int_x^a \mathrm{d}s \sqrt{2(V(s) - E_0)}\right),
\end{equation}
where \( E_0 \) is the ground-state energy and \( a \) is the (positive) classical turning point, defined by \( V(a) = E_0 \).
This implies that, up to subexponential corrections, \( \Delta \propto \Psi_R(0)^2 \propto \Psi_0(0)^2 \).  Upon substitution into Eq.~\eqref{tau-psi}, this leads to the scaling relation~\eqref{tau-delta}.
Therefore, in the assumption that WKB theory holds in a neighborhood of the origin (e.g.\ assuming that the turning points do not approach 0 in the infinite-barrier limit), we see that the DMC tunneling time generically exhibits a \( \Delta^{-1} \) scaling regardless of the specific form of the potential.


\section{Tunneling time in discrete-basis models}
\label{secQIC}
In this section we investigate the tunneling time in PQMC simulations of discrete-basis models, specifically, of quantum spin models. The first model we consider is the quantum Ising chain, described by the following Hamiltonian:
\begin{equation}
\hat{H}=\hat{H}_{\cl}+\hat{H}_{\mathrm{kin}},
\label{H}
\end{equation}
where $\hat{H}_{\cl}=-J\sum_{i=1}^N {\sigma}^{z}_{i} {\sigma}^{z}_{i+1}$ and $\hat{H}_{\mathrm{kin}}=-\Gamma \sum_{i=1}^{N} {\sigma}^{x}_{i}$. $\sigma^x_i$ and $\sigma^z_i$ denote Pauli matrices acting on spins at the lattice site $i$. $N$ is the total number of spins, and we use periodic boundary conditions, i.e., $ {\sigma}^{a}_{N+1}={\sigma}^{a}_{1}$, with $a=x,z$. 
The parameter $J>0$ fixes the strength of the ferromagnetic interactions between nearest-neighbor spins.  In the following, we set $J=1$. $\Gamma$ is the transverse field intensity. Given $\left| x_i \right>$ an eigenstate of ${\sigma^z_i}$ having eigenvalue $x_i=1$ when $\left|x\right>=\left|\uparrow \right>$ and $x_i=-1$ when $\left|x \right>=\left|\downarrow \right>$, the quantum state of $N$ spins is indicated by $\left|\X \right> = \left|x_1 x_2 \dotsc x_N\right>$.
The set $\{|\X \rangle\}$ of $2^N$ states forms the computational basis.

At zero temperature, in the ferromagnetic phase $\Gamma < J$, the quantum Ising chain is characterized by an energy landscape with an effective double-well potential, where the magnetization per spin $M/N$ plays the role of reaction coordinate. The minima of the potential connected by the reaction coordinate correspond to the classical states with magnetization $M \simeq \pm N$. These two states are degenerate when $\Gamma = 0$. For finite $\Gamma$, provided $\Gamma<J$, quantum fluctuations induce tunneling processes between the two minima, lifting the degeneracy in finite systems. 
The energy gap between ground state and first excited state is exponentially small in the system size, i.e., $\Delta \propto \exp\left({-cN}\right)$, where the constant $c$ depends on the transverse field. 
This closing-gap scenario resembles the Landau--Zener avoided level crossings one typically encounters in adiabatic quantum optimization. There, the small gaps are associated to tunneling processes between competing solutions. In order to avoid diabatic transitions to the first excited state, the total annealing time has to scale as $\Delta^{-2}$. These small gaps represent the bottleneck of adiabatic quantum computing, since for hard optimization problems these gaps often close exponentially fast with the system size~\cite{farhi2008,young2008,bapst2013,laumann2015,knysh}.

\subsection{PQMC simulations of discrete-basis models}
Our PQMC simulations for discrete-basis models are based on the continuous-time Green function Monte Carlo algorithm~\cite{SorellaCTGFMC}. This method is exhaustively described in Ref.~\cite{becca_sorella}.  Here, we only sketch the main elements. The simulations with importance sampling are implemented by stochastically evolving the modified imaginary-time Schr\"odinger equation for the product $\rho({\X},t)=\Psi ({\X},t) \Psi_{\gf}({\X})$: 
\begin{equation}
\label{masterfGen}
-\frac{\partial}{\partial t}\rho(\X,t) = \sum_{\X^\prime}  \big[ H_{\X,\X^\prime}-E_{T}\delta_{\X,\X^\prime}  \big] \frac{\Psi_{\gf}({\X})}{\Psi_{\gf}({\X^\prime})} \rho(\X^\prime,t). 
\end{equation}
Here, $\Psi ({\X},t)  \equiv \langle \X|\Psi(t)\rangle$ is the amplitude of the ground-state wave function at imaginary-time $t$, which is assumed to be real and nonnegative. 
The Hamiltonian matrix elements are $H_{\X,\X'}=\langle \X|\hat{H}|\X^\prime\rangle$. $E_{T}$ is again a reference energy used to stabilize the simulations.
When no GWF is used, which corresponds to setting $\Psi_{\gf}({\X})=1$, Eq.~(\ref{masterfGen}) becomes the standard imaginary-time Schr\"odinger equation.
Analogously to the continuous-space simulations described in Sec.~\ref{subs:DMC}, one evolves a population of walkers undergoing spin-flip updates and branching. 
An accurate GWF favors updates toward relevant regions of the configuration space and diminishes walker killings and replications. 
In the infinite imaginary-time limit $t \rightarrow \infty$, attained by iterating many small time steps $ \tau$, the walkers sample spin configurations with a probability distribution proportional to $\Psi_0({\X})\Psi_{\gf}({\X})$, where $\Psi_0({\X})$ is the ground-state wave function.

In this section we consider two types of GWF.
The first is the Boltzmann ansatz:
\beq
\label{boltzmann}
 \Psi_{\gf}{(\X)}=\exp\left[{-\beta E_{\cl} ( {\X} )  }\right].
\eeq 
It resembles the Boltzmann distribution of a classical Ising model with Hamiltonian function $E_{\cl}(\X)=\langle \X|  \hat{H}_{\cl}| \X \rangle$. 
The fictitious temperature $\beta$ is fixed by minimizing the variational energy $ \frac{\langle \Psi_{\gf} |\hat{H}|\Psi_{\gf} \rangle}{\langle \Psi_{\gf} |\Psi_{\gf} \rangle}$ using the stochastic gradient descent method~\cite{SorellaCTGFMC}.
The other GWF is based on a stochastic generative neural network, specifically an unrestricted Boltzmann machine (uRBM)~\cite{inack3}; it is defined as:
\beq \label{ann}
 \Psi_{\gf}(\X)=\sum_{\h} \phi \left(\X,\h \right) \;,
\eeq
where,
\beq
 \phi(\X,\h)=\exp\left[{\sum_{i=1}^{N} \left( K_{1} x_i x_{i+1} +K_{2} h_i h_{i+1}+K_{3}x_i h_i \right) }\right] \;. 	
\label{ann1}
\eeq
The wave function amplitude in each physical, or visible, spin configuration $\X=\left( x_1,x_2,\dots,x_N\right)$ is obtained by integrating over all configurations of the $N$ hidden units $\h=\left( h_1,h_2,\dots,h_N\right)$,  which take the values $h_i= \pm 1$ (with $i=1,\dots,N$). 
Periodic boundary conditions are  considered both in the visible and in the hidden layers, i.e., $x_{N+1}=x_1$ and $h_{N+1}=h_1$. 
The three coupling constants $K_1$, $K_2$, and $K_3$ fix the interaction strengths between nearest-neighbor visible and hidden spins, and between visible--hidden pairs with the same index, respectively. We determine them via variational optimization using the stochastic reconfiguration method.
\begin{figure}
\begin{center}
\includegraphics[width=1.0\columnwidth]{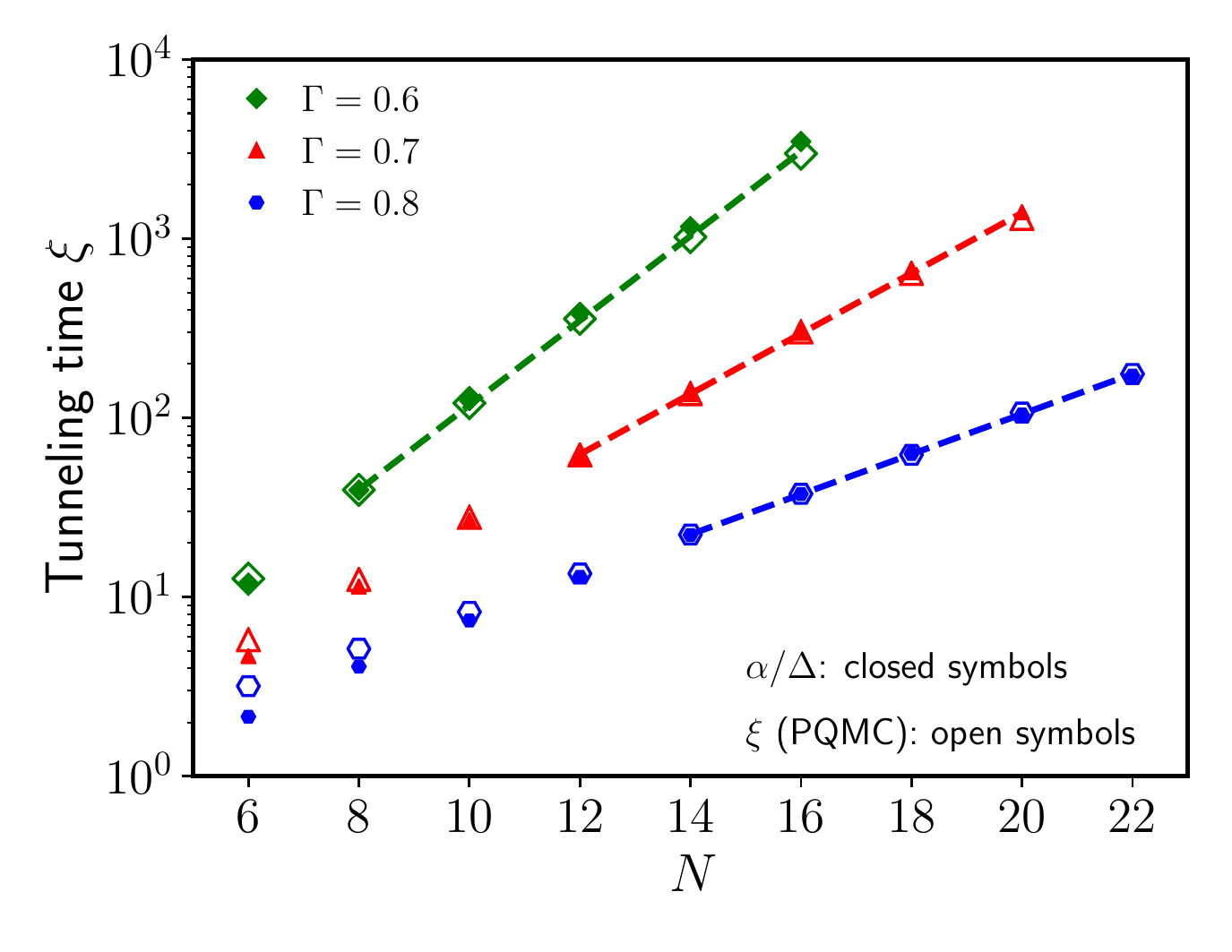}
\caption{(color online). Tunneling time \( \xi \) in PQMC simulations performed with the Boltzmann GWF (open symbols) as a function of the number of spins $N$ in the ferromagnetic Ising chain. Different datasets correspond to different  transverse field intensities $\Gamma$, with the coupling parameter $J=1$. 
The dashed lines represent exponential fitting functions valid in the large-$N$ regime.
The closed symbols represent the inverse gap values $\Delta^{-1}$ computed with the exact formula obtained from the free fermion representation of the quantum Ising chain and rescaled by an appropriate prefactor $\alpha=O(1)$.}
\label{tunqic}
\end{center}
\end{figure}
The analysis reported in Ref.~\cite{inack3} indicated that the optimized uRBMs GWFs are sufficiently accurate to reduce the computational cost of PQMC simulations of the quantum Ising chain down to a polynomial scaling with system size.
Differently from the restricted Boltzmann machine originally introduced as a variational ansatz in Ref.~\cite{carleotroyer}, the uRBM includes intra-layer interactions. In general, this implies that one cannot analytically trace out the hidden spin configurations.\footnote{As shown in Ref.~\cite{collura2019descriptive}, the uRBM can be mapped to a constrained matrix product state. In one dimension, this representation allows for an analytical treatment of the hidden degrees of freedom. However, we aim at a general framework that could be applied irrespectively of the dimensionality and the interaction range.}
In order to use uRBMs as GWFs, we employ the extended PQMC algorithm described in Ref.~\cite{inack3}. It includes a certain number of additional single-spin Metropolis updates of the hidden spins at every PQMC time step. This number has to be made large enough to eliminate spurious correlations among successive walker configuration, which in turn affect the finite-$N_w$ bias.
It is quite important to test if and how the possible residual statistical correlations between successive hidden-spin configurations affect the tunneling dynamics.

\subsection{Tunneling times in quantum Ising chains}
The tunneling time simulations are performed in the ferromagnetic phase, where the quantum Ising chain is characterized by a double-well potential profile.
To measure the PQMC tunneling time \( \xi \), we adopt the protocol of Ref.~\cite{inack2}. 
All walker configurations are initialized with all spin pointing up, corresponding to the classical state with magnetization $M=\sum_i x_i=N$. This state is close to one of the minima of the effective double-well potential. The PQMC simulation is run until $10\%$ of the walkers has crossed the potential barrier, reaching negative magnetization $M<0$. The measurement is repeated about $1000$ times, taking the average and its standard deviation as definition of $\xi$ and of its error bar, respectively.
The PQMC simulations are performed with $N_w = 5000\div10000$, which is found to be sufficient to eliminate any systematic error on $\xi$. 
\begin{figure}
\begin{center}
\includegraphics[width=1.0\columnwidth]{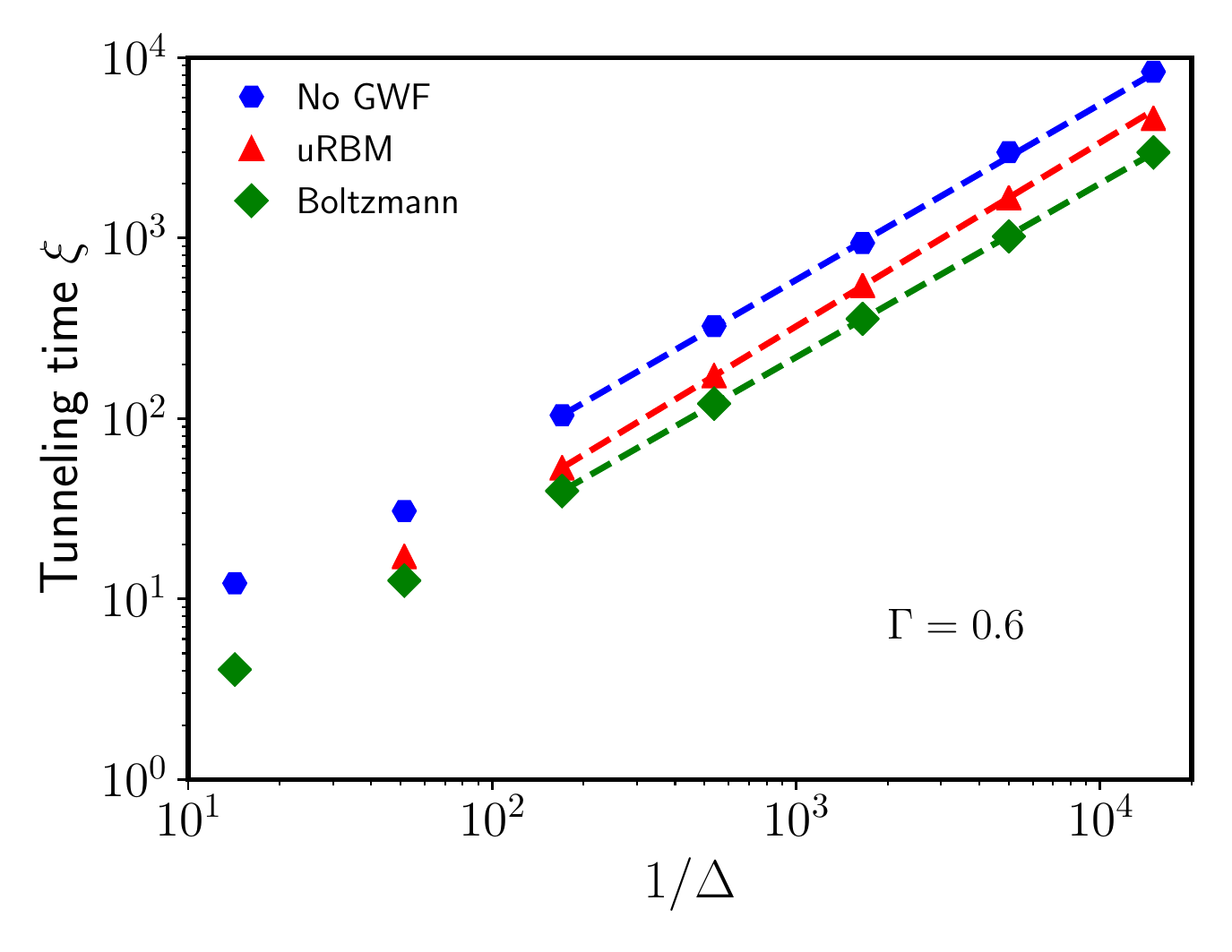}
\caption{(color online). Tunneling times $\xi$ of the PQMC algorithm implemented without GWF (blue pentagons), with the uRBM GWF (red triangles), and with the Boltzmann GWF (green diamonds), as a function of the inverse energy gap $\Delta^{-1}$, for the quantum Ising chain at  $\Gamma=0.6$. 
The dashed lines represent the fitting function $\xi(\Delta)=\alpha\Delta^{-b}$, valid in the large $\Delta^{-1}$ regime. In all three cases, the fitted exponent is $b \simeq 1$ (see Table~\ref{tab:fit2}).
}
\label{compqic}
\end{center}
\end{figure}

\begin{figure}
\begin{center}
\includegraphics[width=0.6\columnwidth]{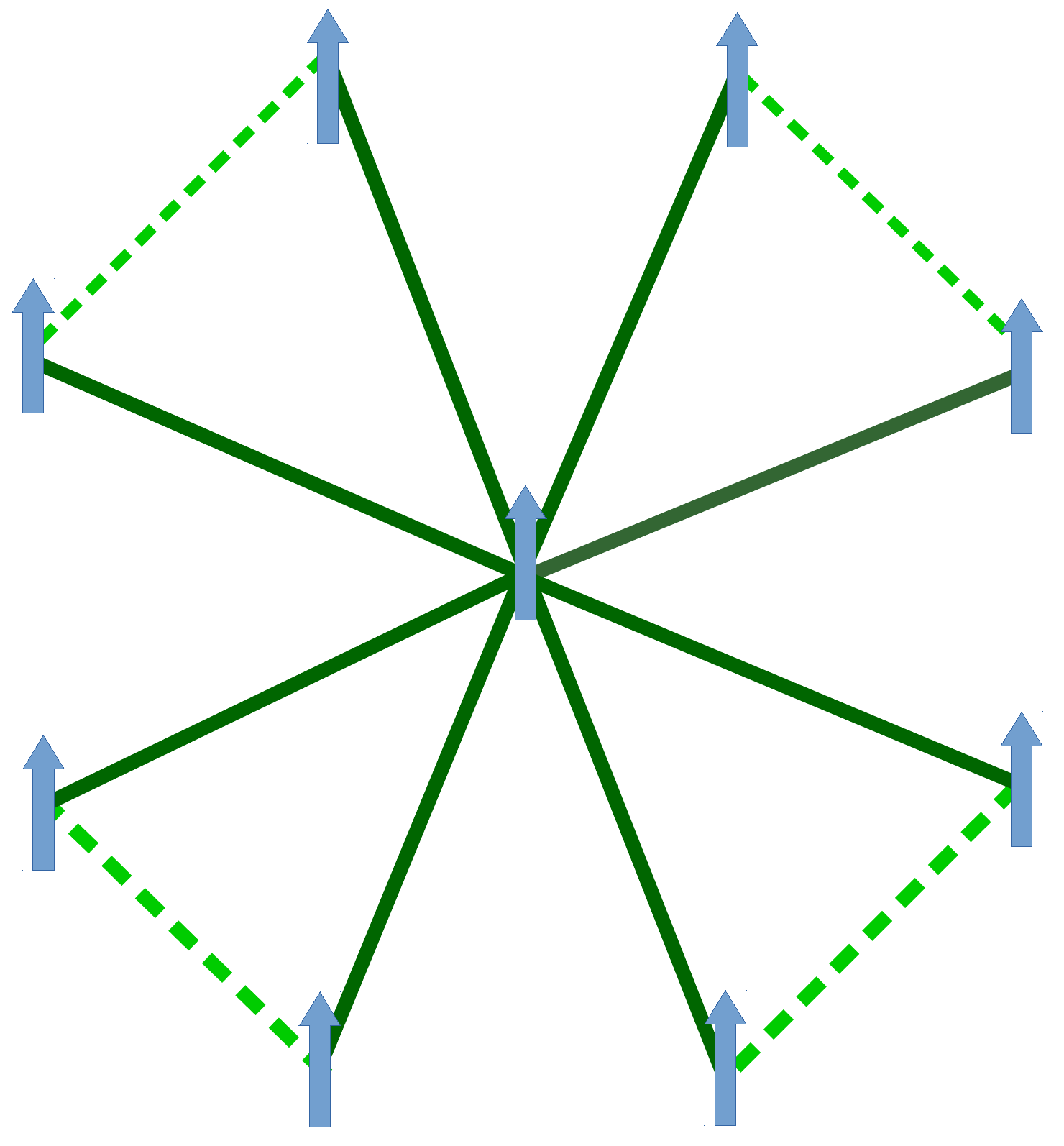}
\caption{(color online). The shamrock, a model of $N$ frustrated spins in a transverse field.  It is made of $K=(N-1)/2$ leaves, each having three spins. The solid dark-green lines represent ferromagnetic interactions (with interaction strength $J$) between the central spin and all the other $N-1$ spins. The dashed light-green lines indicate the antiferromagnetic interactions (with interaction strength $J-\epsilon$) between the outer spins of the same leaf. The overall effect is to create $2^K$ tunneling paths between the degenerate classical ground states.}
\label{shamrock}
\end{center}
\end{figure}

\begin{table}
\begin{tabular}{p{7em}p{5em}l}
PQMC & \( \alpha \) & \( b \)\\\cline{1-3}\\[-2ex]
\textsf{No GWF} & 0.7(2) & 0.97(3) \\
\textsf{uRBM} & 0.32(9) &  1.00(2)\\
\textsf{Boltzmann} & 0.28(5) & 0.96(3)\\
\end{tabular}
\caption{Fitting parameters $\alpha$ and $b$, describing the small-gap behavior of the PQMC tunneling time $\xi$ in the ferromagnetic quantum Ising chain~\eqref{H}, according to the fitting function $ \xi(\Delta) = \alpha \Delta^{-b} $. The error bars also take into account the fluctuations due to choosing different fitting windows.}
\label{tab:fit2}
\end{table}

Figure~\ref{tunqic} displays the tunneling time \( \xi \) obtained with the Boltzmann GWF, as a function of the number of spins $N$, for different transverse field intensities $\Gamma$.
In the large system-size regime, where the energy gap $\Delta$ is small, the exponential growth of \( \xi \) closely matches the scaling of the inverse energy gap $\alpha \Delta^{-1}$, where $\alpha$ is an appropriate prefactor. The energy gap values are computed using the exact formula obtained from the free fermion representation of the quantum Ising chain.
The tunneling times obtained with the Boltzmann GWF and with the uRBM GWF are plotted in Fig.~\ref{compqic} as a function of the corresponding inverse energy gap $\Delta^{-1}$. They are also compared with the results obtained without a GWF (data from Ref.~\cite{inack2}).
In all three cases,  $\xi$ appears to scale asymptotically linearly with the inverse gap. By fitting the three datasets, in the large $\Delta^{-1}$ regime, with the scaling law $\xi(\Delta)=\alpha \Delta^{-b}$, we obtain the values of the fitting parameters $\alpha$ and $b$ reported in Table~\ref{tab:fit2}. In all cases, the exponent is consistent with the linear scaling $b=1$. 

\begin{figure}
\begin{center}
\includegraphics[width=1.0\columnwidth]{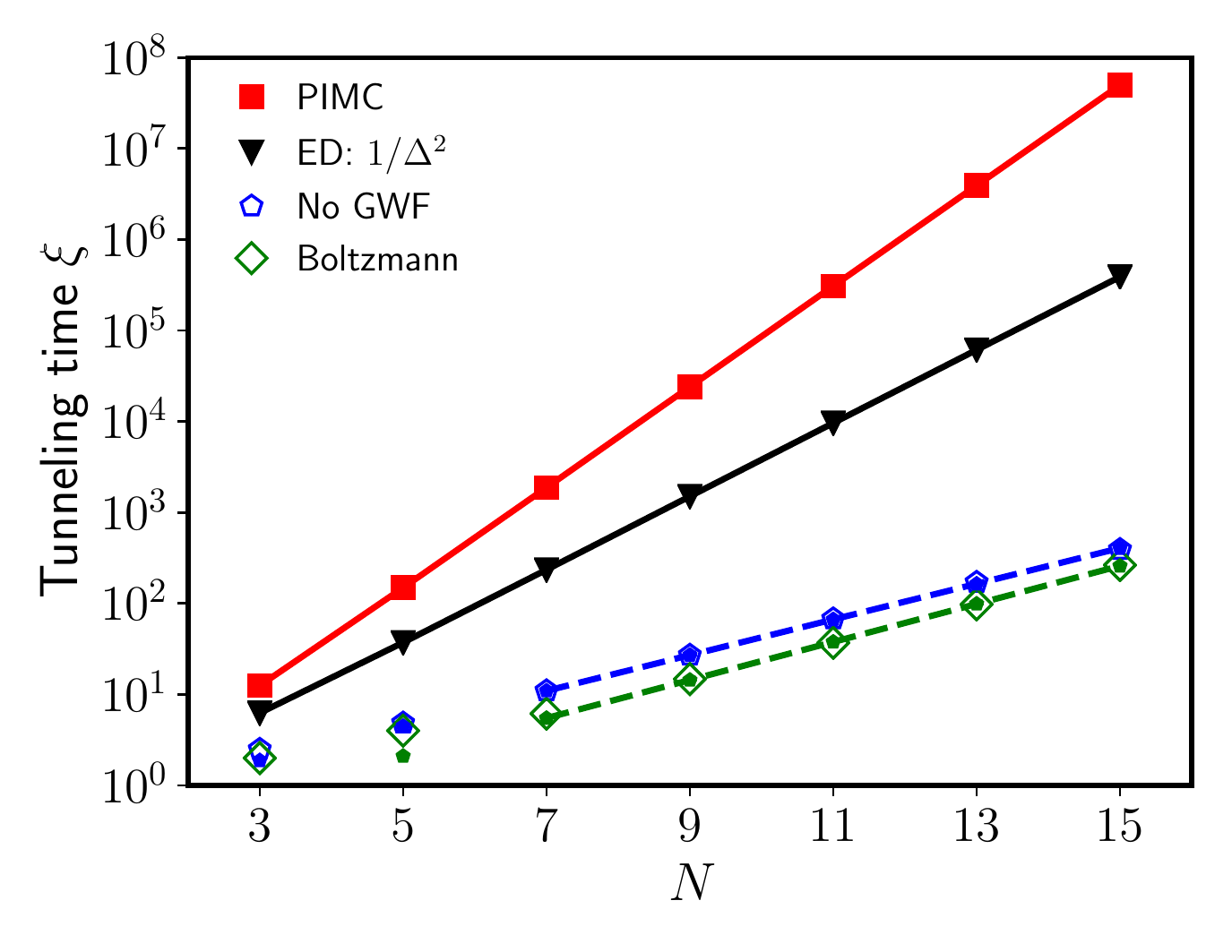}
\caption{(color online). Tunneling time $\xi$ in the shamrock model as a functions of the system size $N$. 
The PQMC results obtained with the Boltzmann GWF (green diamonds) and without GWF (blue pentagons) are compared with the scaling of the incoherent quantum tunneling time $1/\Delta^2$ (black triangles), and with the scaling of the finite-temperature PIMC tunneling time $\xi = 2^K/\Delta^2$~\cite{aminshamrock}, where $K$ is the number of leaves in the shamrock. The values of the gap $\Delta$ are obtained from exact diagonalization.
 The model  parameters are $\Gamma=0.5$, $J=6$, and $\epsilon=0.2$.}
\label{compshamrock}
\end{center}
\end{figure}

\subsubsection*{The shamrock model}
To further study the PQMC tunneling dynamics, we address a more challenging quantum spin Hamiltonian, namely the so-called shamrock model. It is described by the following Hamiltonian:
 \begin{equation}
\hat{H}=-J {\sigma}^{z}_{1} \sum_{i=2}^{N} {\sigma}^{z}_{i}+(J-\epsilon)\sum_{i=1}^{K}  {\sigma}^{z}_{2i} {\sigma}^{z}_{2i+1}  -\Gamma \sum_{i=1}^{N} {\sigma}^{x}_{i}.
\label{Hshamrock}
\end{equation}
The $N$ spins are grouped in $K$ rings, which form the leaves of the shamrock. A central spin interacts with a ferromagnetic coupling $J$ with all the other spins in the system. The outer spins in each ring interact antiferromagnetically with interaction energy \( J-\epsilon \) where $\epsilon\ll J$ indicates a small interaction energy. $\Gamma$ is the intensity of the transverse magnetic field.
The connectivity structure of the shamrock model is visualized in Fig.~\ref{shamrock}.
This model was introduced in Ref.~\cite{aminshamrock} as a paradigmatic case where finite-temperature PIMC algorithms cannot efficiently simulate quantum annealing. This was interpreted as an indication that quantum annealing devices have a high potential to provide a quantum speedup in certain classes of optimization problems. It was indeed shown that in this model the PIMC tunneling time scales as $\xi \propto 2^K \Delta^{-2}$, i.e., exponentially worse than the incoherent quantum tunneling times $ \propto \Delta^{-2}$. This slowdown of the PIMC tunneling dynamics originates from the emergence of multiple homotopy-inequivalent paths for tunneling processes between the competing states.
We measure the PQMC tunneling times in the shamrock model using the Boltzmann GWF. Notice that in this case the classical energy function in Eq.~\eqref{boltzmann} includes the first two terms of the shamrock Hamiltonian~\eqref{Hshamrock}. 
In Fig.~\ref{compshamrock}, these tunneling times are compared with the PQMC results obtained without GWF (data from Ref.~\cite{inack2}), with the scaling of incoherent quantum tunneling, and with the scaling of the PIMC tunneling times. 
In the large $\Delta^{-1}$ regime, the PQMC data with GWF are well described by the fitting function $\xi(\Delta)=\alpha\Delta^{-b}$, where the fitting parameters are $\alpha=0.32(7)$ and $b=1.04(3)$. A similar fit, with $b=0.98(2)$, applies also to the previously reported data, obtained without GWF.
These results indicate that even in the shamrock model the PQMC tunneling times asymptotically scale with the inverse gap, independently on the choice of GWF, confirming the quadratic speedup compared to incoherent quantum tunneling.

\section{Conclusions}
\label{secconclusions}
We have investigated how guiding wave functions affect the tunneling dynamics of PQMC simulations, considering as test beds a continuous-space double-well problem, the ferromagnetic quantum Ising chain, and the shamrock model with frustrated couplings. As GWFs, both approximate variational ansatzes and the numerically computed exact ground-state wave function have been addressed. 
Remarkably, for all GWFs we find a linear relation between tunneling rate and first energy gap in the asymptotic regime of large tunneling time, corresponding to a high potential barrier in the double well, or to large system sizes in the two Ising-type models. 
The semiclassical theory we provided explains this linear relation in the case of double-well--type potentials when the exact ground-state wave function is chosen as GWF.
It is worth stressing that this linear relation represents a quadratic speedup compared to the expected tunneling rate of a physical quantum annealer.
The proof we presented relies on the local validity of the semiclassical approximation for the ground-state wave function. It is an interesting challenge to try to formulate a more general derivation which does not invoke WKB theory, or to instead exhibit a counterexample where a violation of the \( \Delta^{-1} \) scaling may be observed.

Analyzing if and to what extent QMC algorithms can efficiently simulate quantum tunneling is of critical importance to understand if quantum annealing devices can outperform classical optimization methods.
It is well known that accurate PQMC simulations of the equilibrium properties of large-scale systems are only feasible if a sufficiently accurate GWF is used for importance sampling~\cite{foulkes2001quantum}. Indeed, it has been shown that in the quantum Ising chain the GWF can even change the scaling of the computational cost from being exponential to being polynomial in the system size~\cite{inack3}. The results reported here show that an accurate GWF does not alter the quadratic speedup previously reported for PQMC simulations performed without GWF~\cite{inack2}.
Considered together, these findings indicate that PQMC simulations performed with accurate GWFs allow one to efficiently simulate both the equilibrium ground-state properties and also the tunneling dynamics of quantum annealers. Therefore, they can be used as a relevant benchmark in the development of novel quantum annealing devices, and they represent a promising quantum-inspired optimization algorithm~\cite{Microsoft}.
Clearly, more challenging models should be addressed to further benchmark the efficiency of the PQMC tunneling dynamics. Relevant test beds could be Ising spin glasses in higher dimensions. Indeed, one expects that it is harder to obtain accurate variational ansatzes for such models. Suitable candidates are restricted~\cite{carleotroyer,melkorbm2019} and unrestricted Boltzmann machines~\cite{inack3,beachvita2019}. Indeed, both have been shown to be amenable to be used as GWFs in PQMC simulations~\cite{inack3,pilati2019self}. 
Deeper neural network ansatzes, e.g., the deep convolutional neural networks of Ref.~\cite{choo2019study} or the recurrent neural networks of Ref.~\cite{juanreconstruction}, might also be adopted.
We leave these studies for future investigations. \\

\begin{acknowledgments}
\noindent
We acknowledge useful discussions with A.~Gambassi, G.~B.~Mbeng, G.~Santoro, and M.~Troyer.\\
S.P.\ acknowledges financial support from the FAR2018 project of the University of Camerino and from the Italian MIUR under the project PRIN2017 CEnTraL 20172H2SC4.
S.P.\ also acknowledges the CINECA award under the ISCRA initiative, for the availability of high performance computing resources and support. Research at Perimeter Institute is supported in part by the Government of Canada through the Department of Innovation, Science and Economic Development Canada and by the Province of Ontario through the Ministry of Economic Development, Job Creation and Trade.
G.G\ acknowledges support by the ERC under Grant No.~758329 (AGEnTh).
\end{acknowledgments}



%

\end{document}